\documentclass[10pt,aps,floatfix,showpacs]{revtex4}
\usepackage{epsfig}
\usepackage{amsmath}
\usepackage{dcolumn}
\usepackage{bm}
\usepackage{graphicx}
\usepackage{wrapfig}
\usepackage{graphicx}
\usepackage{epstopdf}
\usepackage{epsfig}
\usepackage{IEEEtrantools}
\newcommand{\del}{\delta}

\newcommand{\nn}{\nonumber}

\begin{document}


\title{Late time evolution of the gravitational wave damping in the early Universe}

\author{Gavriil Shchedrin}
\email{shchedrin@nscl.msu.edu}

\affiliation{Department of Physics and Astronomy and National Superconducting Cyclotron Laboratory,
Michigan State University, East Lansing, MI 48824, USA}


\begin{abstract}
An analytical solution for time evolution of the gravitational wave damping
in the early Universe due to freely streaming neutrinos is found in the late time regime.
The solution is represented by a convergent series of spherical Bessel functions of even order
and was possible with the help of a new compact formula for the convolution of spherical Bessel functions of integer order.
\end{abstract}



\maketitle


\section{Introduction}

Thorough analysis of cosmic microwave background (CMB) radiation
provides a unique test of a standard inflationary cosmological
model \cite{infl1,infl2,infl3,infl4,infl5,infl6}. While scalar fluctuations of CMB serve as an invaluable source for
exploring density of matter and radiation and large-scale structure of the universe,
\cite{qm1, qm2, qm3, qm4, qm5, qm6}, observations of tensor fluctuations of CMB open a window for searching after a signature of gravitational waves \cite{gwe1, gwe2, gwe3, gwe4}.

The CMB observations done by Wilkinson Microwave Anisotropy Probe (WMAP) \cite{wmap1}
generally support theoretical predictions based on the standard inflationary cosmological
model. The detailed analysis of the experimental data gives more and
more accurate values \cite{wmap2, wmap3, wmap4} for the most valuable cosmological parameters
such as baryon density, total matter density, Hubble constant, and age of the Universe.

Independently of WMAP measurements,
there is a long quest for a direct observation of cosmological gravitational waves \cite{ligo1}.
Specially designed for this task Laser Interferometer
Gravitational Wave Observatory (LIGO) puts a major effort in this
experimental challenge \cite{ligo2}. A direct observation of cosmological gravitational waves
would serve as a decisive test for validity of the Einstein general theory of relativity
in the same way as the Michelson–-Morley experiment served as a major proof of the Einstein special theory of relativity.

As in the experimental case, cosmological tensor fluctuations
pose a challenge also from a theoretical side \cite{Weincosm1}.
Following S. Weinberg \cite{Weincosm1} we argue that ``The particles of both the cold dark matter and baryonic plasma move too slowly to contribute any anisotropic inertia. In tensor
modes there are no perturbations to densities or streaming velocities, so
there are no perturbations to either the cold dark matter or baryonic plasma
that need to be followed here."

Therefore the only contributions to the anisotropic
inertia tensor are due to photons and neutrinos. Further simplification comes from
the following argument by S. Weinberg \cite{Weincosm1, Weinneutr1, Weinphot1}:
``The anisotropic
inertia tensor is the sum of the contributions from photons and neutrinos,
but photons have a short mean free time before the era of
recombination, and make only a small contribution to the total energy density
afterwards, so their contribution to the anisotropic inertia is small.
This leaves neutrinos (including antineutrinos), which have been traveling
essentially without collisions since the temperature dropped below
about $10^{10}$ K, and which make up a good fraction of the energy density of
the universe until cold dark matter becomes important, at a temperature
about $10^{4}$ K. The tensor part of the anisotropic inertia tensor is given by
\begin{IEEEeqnarray}{l}\label{sw1}
\pi_{ij}(u)=-4\overline{\rho}_{\nu}(u)\int_{0}^{u}K(u-U)h_{ij}'(U)d{U},
\end{IEEEeqnarray}
so the gravitational wave equation
\begin{IEEEeqnarray}{l}\label{sw2}
\frac{d^{2}}{dt^{2}}h_{ij}+3\frac{\dot{a}}{a}\frac{d}{dt}h_{ij}-\frac{\Delta}{a^{2}}h_{ij}=16\pi{G}\pi_{ij},
\end{IEEEeqnarray}
now becomes an integro-differential equation:
\begin{IEEEeqnarray}{l}\label{intdif1}
\frac{d^{2}}{dt^{2}}h_{ij}+3\frac{\dot{a}}{a}\frac{d}{dt}h_{ij}-\frac{\Delta}{a^{2}}h_{ij}=
-64\pi{G}\overline{\rho}_{\nu}(u)\int_{0}^{u}K(u-U)h_{ij}'(U)d{U}."
\end{IEEEeqnarray}
The impact of neutrino source on gravitational wave damping has been thoroughly considered \cite{Weincosm1, Weinneutr1, Weinphot1,wwr1,damp2,damp3,damp4,damp5,damp6,damp7,damp8}.
In this paper we report an analytical solution for the damping of gravitational waves in the early Universe due to freely streaming neutrinos, eq. (\ref{intdif1}), in the late time regime, $u\gg{Q}\gg{1}$.
The solution is represented by an infinite series of spherical Bessel functions of even order. First we shall explain each term in the introduced equations (\ref{sw1}-\ref{intdif1}).

\section{Notations}

We are interested in the time evolution of $h_{ij}(\textbf{x},t)$ that is the tensor perturbation to the metric $g_{\mu\nu}$:
\begin{IEEEeqnarray}{lll}\label{metr1}
g_{00}=-1,\\\nn
g_{i0}=0,\\\nn
g_{ij}(\textbf{x},t)=a^{2}(t)[\del_{ij}+h_{ij}(\textbf{x},t)],
\end{IEEEeqnarray}
where $a(t)$ is the time--dependent Robertson–-Walker scale factor.
The kernel $K(u)$ in eq. (\ref{sw1}) is represented by the sum of three spherical Bessel functions $j_{n}(u)$,
\begin{IEEEeqnarray}{lll}\label{kern1}
K(u)=\frac{1}{15}j_{0}(u)+\frac{2}{21}j_{2}(u)+\frac{1}{35}j_{4}(u).
\end{IEEEeqnarray}
The anisotropic stress tensor $\pi_{ij}(u)$ is obtained from the solution of the Boltzmann equation
for freely streaming neutrinos \cite{Weinneutr1, Weinphot1} which defines the stress–-energy tensor for the Einstein field equations.
The functions $\overline{\rho}_{\nu}(u)$ and $\overline{\rho}_{\gamma}(u)$ give the unperturbed
equilibrium neutrino and photon energy density, correspondingly, which define the ratio:
\begin{IEEEeqnarray}{lll}
f_{\nu}(0)=\frac{\overline{\rho}_{\nu}}{\overline{\rho}_{\nu}+\overline{\rho}_{\gamma}}=
\frac{3\left(\frac{7}{8}\right)\left(\frac{4}{11}\right)^{4/3}}
{1+3\left(\frac{7}{8}\right)\left(\frac{4}{11}\right)^{4/3}}
\simeq{0.40523}.
\end{IEEEeqnarray}
The variable $u$ is the product of
the wave number $k$ and the conformal time,
\begin{IEEEeqnarray}{lll}\label{varu1}
u(t)=k\int_{0}^{t}{\frac{dt'}{a(t')}}.
\end{IEEEeqnarray}
The boundary condition to eq. (\ref{intdif1}) is
\begin{IEEEeqnarray}{lll}
h'_{ij}(0)=0,
\end{IEEEeqnarray}
and it is assumed that we can parameterize the tensor $h_{ij}(u)$ as
\begin{IEEEeqnarray}{lll}
h_{ij}(u)=h_{ij}(0)\chi(u).
\end{IEEEeqnarray}
Introducing the dimensionless quantity
\begin{IEEEeqnarray}{lll}
y\equiv{\frac{a(t)}{a_{eq}}},\\\nn
a_{eq}={a(t=t_{eq})},
\end{IEEEeqnarray}
where $t_{eq}$ is the time of matter-radiation equality,
the general eq. (\ref{intdif1}) can be written as \cite{Weinneutr1}
\begin{IEEEeqnarray}{lll}\label{intdif2}
(1+y)\frac{d^{2}\chi(y)}{dy^{2}}+\left( \frac{2(1+y)}{y}+\frac{1}{2} \right)\frac{d\chi(y)}{dy}+Q^{2}\chi(y)=
-\frac{24 f_{\nu}(0)}{y^{2}}\int_{0}^{y}
K\left(2Q\left(\sqrt{1+y}-\sqrt{1+y'}\right)\right)\frac{d\chi(y')}{dy'}d{y'},
\end{IEEEeqnarray}
with the boundary conditions
\begin{IEEEeqnarray}{lll}\label{bc1}
\chi(0)=1,\\\nn
\left.\frac{d\chi(y)}{d{y}}\right|_{y=0}=0.
\end{IEEEeqnarray}
Eq. (\ref{intdif2}) can be further simplified by the change of variable,
\begin{IEEEeqnarray}{lll}
y=\frac{u(u+4Q)}{4Q^{2}},
\end{IEEEeqnarray}
into
\begin{IEEEeqnarray}{lll}\label{gen1}
\frac{d^{2}\chi(u)}{du^{2}}+\left(\frac{4(u+2Q)}{u(u+4Q)}\right)\frac{d\chi(u)}{du}+\chi(u)=
-{24 f_{\nu}(0)}\left(\frac{4Q}{u(u+4Q)}\right)^{2}\int_{0}^{u}K(u-U)\frac{d\chi(U)}{d{U}}d{U}.
\end{IEEEeqnarray}
Here $Q$ is defined by the ratio of the wave number to its value at the time of matter-radiation equality,
\begin{IEEEeqnarray}{lll}
Q=\sqrt{2}\frac{k}{k_{eq}},\\\nn
k_{eq}=a_{eq}H_{eq},\\\nn
H(t)=\frac{\dot{a}(t)}{a(t)}.
\end{IEEEeqnarray}

\section{Time evolution of the gravitational wave damping}
In the late time regime, $u\gg{Q}\gg{1}$, the general eq. (\ref{gen1}) simplifies into
\begin{IEEEeqnarray}{lll}\label{long1}
\frac{d^{2}\chi(u)}{du^{2}}+\frac{4}{u}\frac{d\chi(u)}{du}+\chi(u)=
\frac{\hat{C}}{u^{4}}\int_{0}^{u}K(u-U)\frac{d\chi(U)}{d{U}}d{U},
\end{IEEEeqnarray}
where $\hat{C}\equiv{-24 f_{\nu}(0)(4Q)^{2}}$.
Below we present the analytical solution for eq. (\ref{long1})
together with boundary conditions (\ref{bc1}).

Clearly we need to find
a specific function that would ``absorb" all derivatives of $u$ and all powers of $u$ on the left hand side of eq. (\ref{long1}). For the differential operator that appears in the left hand side of eq. (\ref{long1}),
\begin{IEEEeqnarray}{lll}\label{op1}
{L}=u^{4}\left(\frac{d^{2}}{du^{2}}+\frac{4}{u}\frac{d}{du}+1\right),
\end{IEEEeqnarray}
these conditions can be satisfied with the function
\begin{IEEEeqnarray}{lll}\label{trial1}
     f_{n}(u)=\frac{n (n+3)}{2n-1}[j_{n-2}(u)+j_{n}(u)]+\frac{(n-2) (n+1)}{2n+3}[j_{n}(u)+j_{n+2}(u)].
\end{IEEEeqnarray}
Applying the differential operator (\ref{op1}) to the function (\ref{trial1}) we obtain a single spherical Bessel function
\begin{IEEEeqnarray}{lll}\label{lhs1}
    {L}[f_{n}(u)]=n(n-2)(n+1)(n+3)(2 n+1)j_{n}(u)
\end{IEEEeqnarray}
which is exactly what we are looking for.
Therefore we can look for the solution of eq. (\ref{long1}) in terms of the expansion:
\begin{IEEEeqnarray}{lll}\label{ex1}
     \chi(u)=\sum_{n=0}^{\infty}c_{n}\left(\frac{n (n+3)}{2n-1}[j_{n-2}(u)+j_{n}(u)]+\frac{(n-2) (n+1)}{2n+3}[j_{n}(u)+j_{n+2}(u)]\right).
\end{IEEEeqnarray}
The left hand side of eq. (\ref{long1}) transforms into
\begin{IEEEeqnarray}{lll}\label{lhs2}
\sum_{n=0}^{\infty}n(n-2)(n+1)(n+3)(2 n+1)c_{n}j_{n}(u).
\end{IEEEeqnarray}
The regular at the origin solution for the homogeneous part of eq. (\ref{long1}),
\begin{IEEEeqnarray}{lll}\label{eqh3}
\chi''_{0}(u)+\frac{4}{u}\chi'_{0}(u)+\chi_{0}(u)=0,
\end{IEEEeqnarray}
is the sum of the two spherical Bessel functions,
\begin{IEEEeqnarray}{lll}\label{longhom1}
\chi_{0}(u)=j_{0}(u)+j_{2}(u).
\end{IEEEeqnarray}
The homogenous part $\chi_{0}(u)$ of the general solution $\chi(u)$ can be already seen as
a linear combination of the first two terms in the expansion ($\ref{ex1}$) for $n=0$ and $n=2$.

\begin{figure}
\centerline{\mbox{\epsfxsize=8cm \epsffile{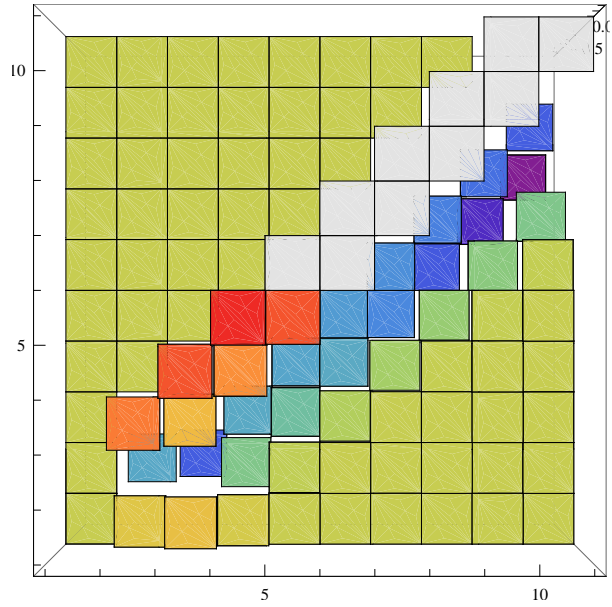}} } \caption{Graphical representation for the
upper but one triangular structure of the matrix $B_{2k,2l}$, eq. (\ref{matr1}).
The matrix indices $l$ and $k$ define a position of the matrix element in the $xy$-plane, while
along $z$-axe we plot its value.}
\label{l1}
\end{figure}

The right hand side of eq. (\ref{long1}) is represented by the convolution of the kernel (\ref{kern1}) with the
first derivative of the unknown function $\chi(u)$ which we are looking for in terms of a series (\ref{ex1}).
Clearly one needs a mathematical tool that relates a convolution of
spherical Bessel functions to a series of those.
In Appendix A we prove a useful formula for the convolution of spherical Bessel functions,
that is not presented in the mathematical literature:
\begin{IEEEeqnarray}{lll}\label{convol1}
\mathcal{J}_{n,m}(u)\equiv{\int_{0}^{u}d{U}j_{m}(u-U)j_{n}(U)}=
\\\nn=\frac{4(-i)^{n+m}}{2i}
    \sum_{l=0}^{\infty}(2l+1)i^{l}j_{l}(u)
 \left(
 \sum_{L=|l-m|\atop {L-n-1\geq{0}}}^{L=l+m}
\frac{\langle{l,0,m,0|L,0\rangle}^2}{L+L^2-n (1+n)}
+\sum_{L=|l-n|\atop {m-L-1\geq{0}}}^{L=l+n}
 \frac{\langle{l,0,n,0|L,0\rangle}^2}{L+L^2-m (1+m)}
\right),
\end{IEEEeqnarray}
where $\langle{l,0,m,0|L,0\rangle}$ are the Clebsch--Gordan coefficients.

The right hand side of eq. (\ref{long1}) is
\begin{IEEEeqnarray}{lll}\label{longser1}
\hat{C}\int_{0}^{u}K(u-U)\frac{d\chi(U)}{d{U}}d{U}=
\hat{C}
\sum_{l=0 \atop {l\in{even}}}^{\infty}j_{l}(u)\sum_{k=0 \atop {k\in{even}}}^{l+2}B_{k,l}c_{k}.
\end{IEEEeqnarray}
Here the matrix $B_{k,l}$ is generated by
the convolution of the first derivative of the function $\chi(u)$,
\begin{IEEEeqnarray}{lll}\label{der1}
     \chi'(u)=\sum_{n}c_{n}\left(
     j_{n-3}(u)\frac{(-2+n) n (3+n)}{(-3+2 n) (-1+2 n)}
     +j_{n-1}(u)\frac{n \left(-3-4 n+n^2\right)}{(-3+2 n) (3+2 n)}+\right.\\\nn
     \left.
     j_{n+1}(u)\frac{(-1-n) \left(2+6 n+n^2\right)}{(-1+2 n) (5+2 n)}+
     j_{n+3}(u)\frac{(-1-n) (-2+n) (3+n)}{(3+2 n) (5+2 n)}
     \right),
\end{IEEEeqnarray}
with the kernel (\ref{kern1}) by means of eq. (\ref{convol1}),
and for integer $k\in [0,10]$ and $l\in [0,10]$ is
\begin{IEEEeqnarray}{lll}\label{matr1}
B_{2k,2l}\equiv{}
\left(
\begin{array}{ccccccccccc}
 0 & \frac{1}{15} & \frac{1}{10} & \frac{13}{350} & \frac{17}{4900} & -\frac{1}{3780} & \frac{1}{19404} & -\frac{29}{1981980} & \frac{1}{193050} & -\frac{37}{17381650} & \frac{41}{42031990} \\\\
 0 & -\frac{1}{3} & -\frac{1}{2} & -\frac{13}{70} & -\frac{17}{980} & \frac{1}{756} & -\frac{5}{19404} & \frac{29}{396396} & -\frac{1}{38610} & \frac{37}{3476330} & -\frac{41}{8406398} \\\\
 0 & \frac{4}{15} & \frac{7}{55} & -\frac{5671}{17325} & -\frac{10523}{44100} & -\frac{733}{13860} & -\frac{1453}{640332} & -\frac{29}{77297220} & \frac{173}{5855850} & -\frac{17279}{1095043950} & \frac{65887}{8322334020} \\\\
 0 & 0 & \frac{18}{55} & \frac{793}{3465} & -\frac{3013}{8820} & -\frac{13507}{41580} & -\frac{2077}{23716} & -\frac{1247}{220220} & \frac{353}{1351350} & -\frac{1147}{56156100} & -\frac{8077}{3627684060} \\\\
 0 & 0 & 0 & \frac{88}{225} & \frac{68323}{209475} & -\frac{6611}{17955} & -\frac{1343}{3234} & -\frac{3652637}{30060030} & -\frac{4375681}{491891400} & \frac{5229247}{10220410200} & -\frac{245303}{3530687160} \\\\
 0 & 0 & 0 & 0 & \frac{26}{57} & \frac{1661}{3933} & -\frac{81703}{204930} & -\frac{4520839}{8918910} & -\frac{39881}{257400} & -\frac{67721729}{5631654600} & \frac{174227819}{231512200920} \\\\
 0 & 0 & 0 & 0 & 0 & \frac{12}{23} & \frac{74315}{143451} & -\frac{244553}{567567} & -\frac{302161}{504504} & -\frac{62891453}{334486152} & -\frac{95878295}{6355236888} \\\\
 0 & 0 & 0 & 0 & 0 & 0 & \frac{238}{405} & \frac{701017}{1142505} & -\frac{2355893}{5077800} & -\frac{5773813}{8353800} & -\frac{7011533}{31744440} \\\\
 0 & 0 & 0 & 0 & 0 & 0 & 0 & \frac{304}{465} & \frac{11539}{16275} & -\frac{119899}{240975} & -\frac{1434877}{1831410} \\\\
 0 & 0 & 0 & 0 & 0 & 0 & 0 & 0 & \frac{18}{25} & \frac{39997}{49725} & -\frac{1405973}{2645370} \\\\
 0 & 0 & 0 & 0 & 0 & 0 & 0 & 0 & 0 & \frac{92}{117} & \frac{602003}{669123}
\end{array}
\right)
\end{IEEEeqnarray}

We should notice an unpleasant feature of the matrix of coefficients (\ref{matr1}): the first row up to a factor $k=-{5}$ is identical to the second row.
This is a direct response to the symmetry of the introduced function $f_{n}(u)$, eq. (\ref{trial1}).
The functions $f_{n}(u)$ for $n=0$ and $n=2$ are exactly the same up to the factor $k=-{5}$,
\begin{IEEEeqnarray}{lll}\label{trialf}
     f_{0}(u)= -\frac{2}{3}(j_{0}(u)+j_{2}(u)),\\\nn
     f_{2}(u)= \frac{10}{3}(j_{0}(u)+j_{2}(u)).
\end{IEEEeqnarray}
Therefore the rank of the matrix (\ref{matr1}) is $\verb"Rank"[B]=N-1$.
This is a real obstacle because it leads to inconsistency with the boundary conditions. Indeed, the boundary conditions (\ref{bc1}) are met if we set
\begin{IEEEeqnarray}{lll}
-\frac{2}{3}c_{0}+\frac{10}{3}c_{2}=1.
\end{IEEEeqnarray}
On the other hand, the linear dependence of the matrix is equivalent to
\begin{IEEEeqnarray}{lll}
-\frac{2}{3}c_{0}+\frac{10}{3}c_{2}=0.
\end{IEEEeqnarray}
In order to avoid this unpleasant feature we can start summation in the series (\ref{ex1})
from $n=2$ instead of $n=0$ which is equivalent to setting
\begin{IEEEeqnarray}{lll}
c_{0}\equiv{0},
\end{IEEEeqnarray}
and thus the boundary conditions (\ref{bc1}) are met if we set
\begin{IEEEeqnarray}{lll}
c_{2}=\frac{3}{10}.
\end{IEEEeqnarray}
Absence of the $j_{2}(u)$ term in the left hand side of eq. (\ref{lhs2})
leads to a restriction on the first coefficients in eq. (\ref{longser1}):
\begin{IEEEeqnarray}{lll}
-\frac{1}{3} c_2+\frac{4}{15} c_4=0.
\end{IEEEeqnarray}
Finally we get the system of linear equations
\begin{IEEEeqnarray}{lll}
n(n-2)(n+1)(n+3)(2 n+1)c_{n}=\hat{C}\sum_{k=0\atop{k\in{even}}}^{n+2}B_{k,n}c_{k}
\end{IEEEeqnarray}
that returns solution (for even $n$ and $k$)
\begin{IEEEeqnarray}{lll}\label{longsol1}
c_{n+2}=\frac{n(n-2)(n+1)(n+3)(2 n+1)c_{n}-\hat{C}\sum_{k=0\atop{k\in{even}}}^{n}B_{k,n}c_{k}}{\hat{C}B_{n+2,n}}.
\end{IEEEeqnarray}

In the limit $Q\gg{1}$ and owing to $\hat{C}\equiv{-24 f_{\nu}(0)(4Q)^{2}}$ we have $Q$-independent solution:
\begin{IEEEeqnarray}{lll}\label{longsol1}
c_{n+2}=-\frac{\sum_{k=0\atop{k\in{even}}}^{n}B_{k,n}c_{k}}{B_{n+2,n}},
\end{IEEEeqnarray}
and therefore for $n\in{[0,9]}$ we have
\begin{eqnarray}
c_{2n}=\left\{0,\frac{3}{10},\frac{3}{8},\frac{5}{16},\frac{35}{128},\frac{63}{256},\frac{231}{1024},\frac{429}{2048},\frac{6435}{32768},\frac{12155}{65536}\right\}
\end{eqnarray}
which completes our series solution eq. (\ref{ex1}).

\begin{figure}
\centerline{\mbox{\epsfxsize=10cm \epsffile{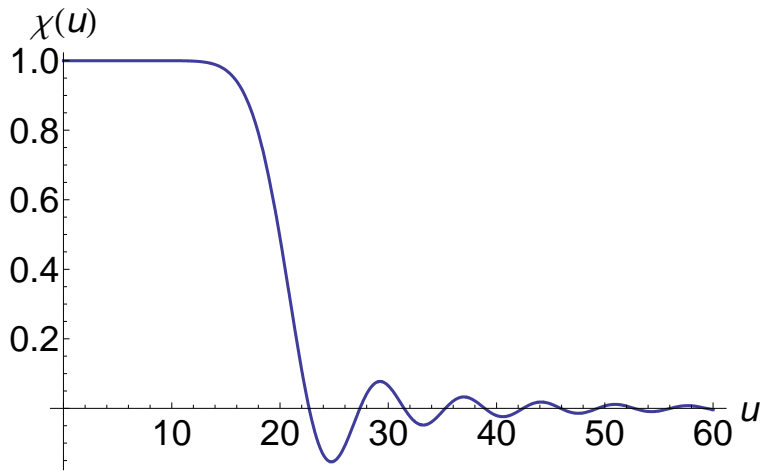}} } \caption{Late time evolution of the gravitational wave damping $\chi(u)$ in the early Universe}
\label{l1}
\end{figure}

\section{Conclusion}

We have analyzed the problem of gravitational wave damping 
in the early Universe due to freely streaming neutrinos in the late time regime $u\gg{Q}\gg{1}$.
As in the opposite limit $u\ll{Q}$ \cite{wwr1}, the solution is represented by a convergent series of spherical Bessel functions of even order and is independent of the $Q$-value. Thus we conclude that the problem 
gravitational wave damping in the early Universe due to freely streaming neutrinos is completely solved in 
an analytical way in both early and late time limits.

\section*{Acknowledgements}
I am thankful to W.W. Repko and V.G. Zelevinsky for an encouragement and many helpful suggestions. I am indebted to N.O. Birge, J.T. Linnemann, C. Schmidt, M. Shapiro, and A.L. Volberg for valuable remarks.

\section*{Appendix A}
Here we derive the convolution integral (\ref{convol1}) of spherical Bessel functions for integer orders $n,m$.
Theoretical efforts for such a convolution integral were put forward in \cite{wwr1} but here we report
a new compact formula which has a clear exchange symmetry ($n{\longleftrightarrow}m$)
and can be readily applied for a practical calculations.

Starting with the integral
\begin{IEEEeqnarray}{lll}\label{start1}
    \mathcal{J}_{n,m}(u)\equiv{\int_{0}^{u}d{U}j_{m}(u-U)j_{n}(U)}
\end{IEEEeqnarray}
we prove (\ref{convol1}).
First, we represent a spherical Bessel function as
a Fourier transformation of the Legendre polynomial $P_{n}(z)$,
\begin{IEEEeqnarray}{lll}\label{legendr1}
    j_{n}(u)=\frac{(-i)^{n}}{2}\int_{-1}^{1}d{s}\exp(ius)P_{n}(s).
\end{IEEEeqnarray}
Substitution of (\ref{legendr1}) into (\ref{start1}) leads to
\begin{IEEEeqnarray}{lll}\label{a3}
    \mathcal{J}_{n,m}(u)=\frac{(-i)^{n+m}}{4}\int_{-1}^{1}d{s}\int_{-1}^{1}d{t}\exp(iut)P_{m}(s)P_{n}(t)
    \int_{0}^{u}d{U}\exp(iUs-iUt)=\\\nn=
    \frac{(-i)^{n+m}}{4i}
    \int_{-1}^{1}d{s}\int_{-1}^{1}d{t}P_{m}(s)P_{n}(t)\frac{\exp(ius)-\exp(iut)}{(s-t)}.
\end{IEEEeqnarray}
Now we employ the Legendre function of the second kind $Q_{n}(z)$ defined as
\begin{IEEEeqnarray}{lll}\label{a4}
    Q_{n}(z)=\frac{1}{2}\int_{-1}^{1}d{z'}\frac{P_{n}(z')}{z-z'}.
\end{IEEEeqnarray}
Performing the integrations over $t$ and $s$ we obtain
\begin{IEEEeqnarray}{lll}\label{a5}
    \frac{(-i)^{n+m}}{4i}
    \int_{-1}^{1}d{s}\int_{-1}^{1}d{t}P_{m}(s)P_{n}(t)\frac{\exp(ius)-\exp(iut)}{(s-t)}=\\\nn=
    \frac{(-i)^{n+m}}{2i}\int_{-1}^{1}d{t}\exp(iut)
    \left[P_{n}(t)Q_{m}(t)+P_{m}(t)Q_{n}(t)\right].
\end{IEEEeqnarray}
Further we reincarnate spherical Bessel functions by decomposing plane waves in terms of Legendre polynomials:
\begin{IEEEeqnarray}{lll}\label{a6}
    \exp(iut)=\sum_{l=0}^{\infty}(2l+1)i^{l}j_{l}(u)P_{l}(t),
\end{IEEEeqnarray}
which leads to
\begin{IEEEeqnarray}{lll}\label{a7}
    \frac{(-i)^{n+m}}{2i}\int_{-1}^{1}d{t}\exp(iut)
    \left[P_{n}(t)Q_{m}(t)+P_{m}(t)Q_{n}(t)\right]=\\\nn=
    \frac{(-i)^{n+m}}{2i}
    \sum_{l=0}^{\infty}(2l+1)(-i)^{-l}j_{l}(u)
    \int_{-1}^{1}d{t}P_{l}(t)
    \left[P_{n}(t)Q_{m}(t)+P_{m}(t)Q_{n}(t)\right].
\end{IEEEeqnarray}
The angular momentum coupling simplifies the product of Legendre polynomials,
\begin{IEEEeqnarray}{lll}\label{a8}
    P_{l}(x)P_{m}(x)=\sum_{L=|l-m|}^{L=l+m}\langle{l,0,m,0|L,0\rangle}^2P_{L}(x),
\end{IEEEeqnarray}
in terms of the Clebsch–-Gordan coefficients $\langle{l,0,m,0|L,0\rangle}$.
Introducing
\begin{IEEEeqnarray}{lll}\label{a10}
    W_{m-1}(z)=2\sum_{k=0}^{[\frac{m-1}{2}]}\frac{(m-2k-1)}{(2k+1)(m-k)}P_{m-2k-1}(z),
\end{IEEEeqnarray}
and using the analog of eq. (\ref{a8}),
\begin{IEEEeqnarray}{lll}\label{a12}
 P_{l}(z)Q_{m}(z)=\sum_{L=|l-m|}^{L=l+m}\langle{l,0,m,0|L,0\rangle}^2
 \left(
 Q_{L}(z)+W_{L-1}(z)\right)
 -P_{l}(z)W_{m-1}(z),
\end{IEEEeqnarray}
we come to
\begin{IEEEeqnarray}{llll}\label{a11}
    \int_{-1}^{1}d{t}P_{l}(t)
    \left[P_{n}(t)Q_{m}(t)+P_{m}(t)Q_{n}(t)\right]=\\\nn=
    \int_{-1}^{1}d{t}
    \left[P_{n}(t)\underbrace{P_{l}(t)Q_{m}(t)}_{I}+\underbrace{P_{l}(t)P_{m}(t)}_{II}Q_{n}(t)\right]=\\\nn=
    \int_{-1}^{1}d{t}
    \left[
    P_{n}(t)\left(\sum_{L=|l-m|}^{L=l+m}\langle{l,0,m,0|L,0\rangle}^2
     \left(
     Q_{L}(t)+W_{L-1}(t)\right)-P_{l}(t)W_{m-1}(t)\right)\right.\\\nn+
     \left.\sum_{L=|l-m|}^{L=l+m}\langle{l,0,m,0|L,0\rangle}^2
       P_{L}(t)Q_{n}(t)
    \right],
\end{IEEEeqnarray}
where in the first term we have decomposed the product $P_{l}(t)Q_{m}(t)$, whereas in the second term
we decomposed the product $P_{l}(t)P_{m}(t)$. Using the parity identity,
\begin{IEEEeqnarray}{lll}\label{b1}
    \int_{-1}^{1}d{t} P_{n}(t)Q_{L}(t)=- \int_{-1}^{1}d{t}P_{L}(t)Q_{n}(t),
\end{IEEEeqnarray}
we obtain
\begin{IEEEeqnarray}{lll}
    \int_{-1}^{1}d{t}P_{l}(t)
    \left[P_{n}(t)Q_{m}(t)+P_{m}(t)Q_{n}(t)\right]=\\\nn=
    \int_{-1}^{1}d{t}
    \left[
    P_{n}(t)\left(\sum_{L=|l-m|}^{L=l+m}\langle{l,0,m,0|L,0\rangle}^2
     W_{L-1}(t)-P_{l}(t)W_{m-1}(t)\right)\right].
    \end{IEEEeqnarray}
Therefore
\begin{IEEEeqnarray}{lll}
\int_{-1}^{1}dt\sum_{L=|l-m|}^{L=l+m}\langle{l,0,m,0|L,0\rangle}^2P_{n}(t)W_{L-1}(t)=\\\nn=
\int_{-1}^{1}dt\sum_{L=|l-m|}^{L=l+m}\langle{l,0,m,0|L,0\rangle}^2
\sum_{k=0}^{[(L-1)/2]}\frac{2L-4k-1}{(2k+1)(L-k)}P_{L-2k-1}(t)P_{n}(t)=
4\sum_{L=|l-m|\atop {L-n-1\geq{0}}}^{L=l+m}
\frac{\langle{l,0,m,0|L,0\rangle}^2}{L+L^2-n (1+n)}.
\end{IEEEeqnarray}
On the other hand,
\begin{IEEEeqnarray}{lll}
 \int_{-1}^{1}dtP_{l}(t)W_{m-1}(t)P_{n}(t)=\\\nn=
\int_{-1}^{1}dt\sum_{k=0}^{[(m-1)/2]}\frac{2m-4k-1}{(2k+1)(m-k)}P_{m-2k-1}(t)P_{l}(t)P_{n}(t)=\\\nn=
\int_{-1}^{1}dt\sum_{k=0}^{[(m-1)/2]}\frac{2m-4k-1}{(2k+1)(m-k)}P_{m-2k-1}(t)
 \sum_{L=|l-n|}^{L=l+n}\langle{l,0,n,0|L,0\rangle}^2P_{L}(t)=
-4\sum_{L=|l-n|\atop {m-L-1\geq{0}}}^{L=l+n}
 \frac{\langle{l,0,n,0|L,0\rangle}^2}{L+L^2-m (1+m)}.
\end{IEEEeqnarray}
Thus, the identity (\ref{convol1}) is demonstrated,
\begin{IEEEeqnarray}{lll}
    \int_{-1}^{1}d{t}
    \sum_{L=|l-m|}^{L=l+m}\langle{l,0,m,0|L,0\rangle}^2
    P_{n}(t)W_{L-1}(t)-
    \int_{-1}^{1}d{t}P_{n}(t)P_{l}(t)W_{m-1}(t)=\\\nn=
 4\sum_{L=|l-m|\atop {L-n-1\geq{0}}}^{L=l+m}
\frac{\langle{l,0,m,0|L,0\rangle}^2}{L+L^2-n (1+n)}
+4\sum_{L=|l-n|\atop {m-L-1\geq{0}}}^{L=l+n}
 \frac{\langle{l,0,n,0|L,0\rangle}^2}{L+L^2-m (1+m)}.
\end{IEEEeqnarray}

\end{document}